%% file: main.tex
\documentclass[11pt,twoside]{article}
\usepackage{ACL2023}
\usepackage{times}
\usepackage{xspace}
\usepackage{enumitem}
\usepackage{multirow}
\usepackage{booktabs} 
\usepackage{color} 
\usepackage{graphicx}
\usepackage{pifont}
\usepackage{tabularx}
\usepackage{listings}
\usepackage{setspace}
\usepackage{colortbl}
\usepackage[T2A,T1]{fontenc}
\usepackage[utf8]{inputenc}
\usepackage[russian,english]{babel}
\usepackage{xcolor}
\usepackage{tikz}
\usepackage{times}
\usetikzlibrary{positioning, backgrounds}
\usepackage{standalone}
\usepackage{url}  

\usepackage[T1]{fontenc}
\usepackage[utf8]{inputenc}
\usepackage{microtype}
\usepackage{inconsolata}

\newcolumntype{R}[2]{%
    >{\adjustbox{angle=#1,lap=\width-(#2)}\bgroup}%
    l%
    <{\egroup}%
}

\usepackage{fancyhdr} 
\title{\extremebb: A Database for Large-Scale Research into Online Hate, Harassment, the Manosphere and Extremism}

\author{Anh V. Vu$^{\dagger}$, Lydia Wilson$^{\dagger}$, Yi Ting Chua$^{\star}$, Ilia Shumailov$^{\diamond}$, Ross Anderson$^{\dagger\ddagger}$\\
  $^{\dagger}$University of Cambridge, $^{\star}$University of Alabama\\
$^{\diamond}$University of Oxford, $^{\ddagger}$University of Edinburgh
}

\begin{document}
\include{configs}
\include{stats}

\maketitle
\begin{abstract}
We introduce \extremebb, a textual database of over \nPostsTotalShorten~posts made by \nUsersTotalShorten~users on 12 extremist bulletin board forums promoting online hate, harassment, the manosphere and other forms of extremism. It enables large-scale analyses of qualitative and quantitative historical trends going back two decades: measuring hate speech and toxicity; tracing the evolution of different strands of extremist ideology; tracking the relationships between online subcultures, extremist behaviours, and real-world violence; and monitoring extremist communities in near real time. This can shed light not only on the spread of problematic ideologies but also the effectiveness of interventions. \extremebb comes with a robust ethical data-sharing regime that allows us to share data with academics worldwide. Since 2020, access has been granted to 49 licensees in 16 research groups from 12 institutions.
\end{abstract}

\begin{table*}[t]
\centering
\caption{The category, abbreviation, number of users, boards, threads, posts, and collection period of the 12 forums.}
\small
\setlength{\tabcolsep}{.27em}
\begin{tabular}{lrrrrrrrr}
\toprule
\textbf{Category} & \textbf{Forum Name} & \textbf{Abbr.}  & \textbf{Users} & \textbf{Boards} & \textbf{Threads} & \textbf{Posts (\% Empty)} & \textbf{Data Period}\\
\midrule
\arrayrulecolor{black!20}
{\scshape White} & \stormfront & \stabbr & \nUsersStormfront & \nBoardsStormfront & \nThreadsStormfront & \nPostsStormfront~(\nEmptyPostsPropsStormfront\%) & \nDataPeriodStormfront \\
{\scshape Supremacy} & \vnn & \vaabbr & \nUsersVanguardNewsNetwork & \nBoardsVanguardNewsNetwork & \nThreadsVanguardNewsNetwork & \nPostsVanguardNewsNetwork~(\nEmptyPostsPropsVanguardNewsNetwork\%) & \nDataPeriodVanguardNewsNetwork \\
\midrule
\multirow{2}{*}{\iccat} & \incelsis & \iiabbr & \nUsersIncelsis & \nBoardsIncelsis & \nThreadsIncelsis & \nPostsIncelsis~(\nEmptyPostsPropsIncelsis\%) & \nDataPeriodIncelsis \\
& \incelsnet & \inabbr & \nUsersIncelsnet & \nBoardsIncelsnet & \nThreadsIncelsnet & \nPostsIncelsnet~(\nEmptyPostsPropsIncelsnet\%) & \nDataPeriodIncelsnet \\
\midrule
\multirow{2}{*}{\lkcat} & \lookism & \lkabbr & \nUsersLookism & \nBoardsLookism & \nThreadsLookism & \nPostsLookism~(\nEmptyPostsPropsLookism\%) & \nDataPeriodLookism \\
& \looksmax & \lsabbr & \nUsersLooksmax & \nBoardsLooksmax & \nThreadsLooksmax & \nPostsLooksmax~(\nEmptyPostsPropsLooksmax\%) & \nDataPeriodLooksmax \\
\midrule
{\scshape Pickup} & \rooshv & \rvabbr & \nUsersRooshV & \nBoardsRooshV & \nThreadsRooshV & \nPostsRooshV~(\nEmptyPostsPropsRooshV\%) & \nDataPeriodRooshV \\
{\scshape Artistry} & \pa & \paabbr & \nUsersPickupArtist & \nBoardsPickupArtist & \nThreadsPickupArtist & \nPostsPickupArtist~(\nEmptyPostsPropsPickupArtist\%) & \nDataPeriodPickupArtist \\
\midrule
{\scshape Men's} & \mgtow & \mgabbr & \nUsersMenGoingTheirOwnWay & \nBoardsMenGoingTheirOwnWay & \nThreadsMenGoingTheirOwnWay & \nPostsMenGoingTheirOwnWay~(\nEmptyPostsPropsMenGoingTheirOwnWay\%) & \nDataPeriodMenGoingTheirOwnWay \\
{\scshape Movement} & \gyow & \gyabbr & \nUsersGoingYourOwnWay & \nBoardsGoingYourOwnWay & \nThreadsGoingYourOwnWay & \nPostsGoingYourOwnWay~(\nEmptyPostsPropsGoingYourOwnWay\%) & \nDataPeriodGoingYourOwnWay \\
\midrule
{\scshape Online} & \kiwifarms & \kfabbr & \nUsersKiwiFarms & \nBoardsKiwiFarms & \nThreadsKiwiFarms & \nPostsKiwiFarms~(\nEmptyPostsPropsKiwiFarms\%) & \nDataPeriodKiwiFarms \\
{\scshape Harassment} & \lolcow & \lcabbr & \nUsersLolcow & \nBoardsLolcow & \nThreadsLolcow & \nPostsLolcow~(\nEmptyPostsPropsLolcow\%) & \nDataPeriodLolcow \\
\arrayrulecolor{black}
\midrule
6 Categories & 12 Forums & & \nUsersTotal & \nBoardsTotal & \nThreadsTotal & \nPostsTotal~(\nEmptyPostsPropsTotal\%) & \nDataPeriodTotal \\
\bottomrule
\end{tabular}
\label{tab:database-taxonomy}
\end{table*}

\section{Online Extremist Forums}
Online hate, harassment, the manosphere and extremism are growing and pernicious problems, associated with real-world threats and violent crime~\cite{aghazadeh2018gamergate, onlineextremism,vu2023no}. Large-scale quantitative studies of these emergent problems and the links between them have been limited by the difficulties of collecting high-quality data, and of sharing these data ethically with other academic researchers. The recent withdrawal of free public access to Twitter is a real challenge~\cite{twitterapinotfree}. Extremist material is concentrated in a variety of other channels and places online, particularly on Reddit and in forums. 

While Reddit data has been collected and shared widely~\cite{baumgartner2020pushshift}, the research community has lacked a comprehensive collection of other online forum data at scale. Forums are structured into themed boards, each a collection of threads with one or more posts. Posts have not just textual content but valuable metadata such as posting time, authors, their joining date and reputation. 

We collect online forum data and make it available to researchers. Our collection methodology is dynamic. We first compiled a list of known online communities using in-house and external experts, and shortlisted publicly accessible active extremist forums with at least 50k posts. As some forums are agile in the sense that they are short-lived or their domains rotate, a feedback loop was created with experts monitoring the list and alerting collection engineers of any changes. We also actively monitor extremist Telegram and Discord channels to spot new forums. At present, most \extremebb forums originate from the USA, grouped into the following six categories based on the nature of their content and their self-definition. 

\para{\wscat} This contains the two largest and longest-running race-hate, antisemitic and white supremacist forums. Some forum users were reportedly involved in serious crime, and these forums are associated with terrorist atrocities such as the Wisconsin Sikh temple shooting, the Pittsburgh police officer shooting, and the Norway attacks. Our collection is significantly larger than those reported in previous work~\cite{holt2020examining,kleinberg2021temporal}.

\para{\iccat} This is an extreme misogynistic ideology whose dominant theme is its members' inability to have sexual relationships with women. Forum users envy attractive men and blame feminism and female nature for their plight, indulging in violent fantasies and hate speech. The two major forums we collect are larger than datasets of incel material used in existing work~\cite{baele2019incel,jaki2019online,o2022political}.

\para{\lkcat} This consists of the two largest forums on techniques to enhance male physical beauty, ranging from non-invasive methods such as styling and workouts to extreme ones involving cosmetic surgery, synthetic hormones, and skin-lightening products. This is strongly linked to the incel subcultures, with users often entangled with misogyny involving a stereotype of female superficiality that requires men to invest in a masculine appearance.

\begin{figure}[t]
    \centering
    \includegraphics[width=.47\textwidth]{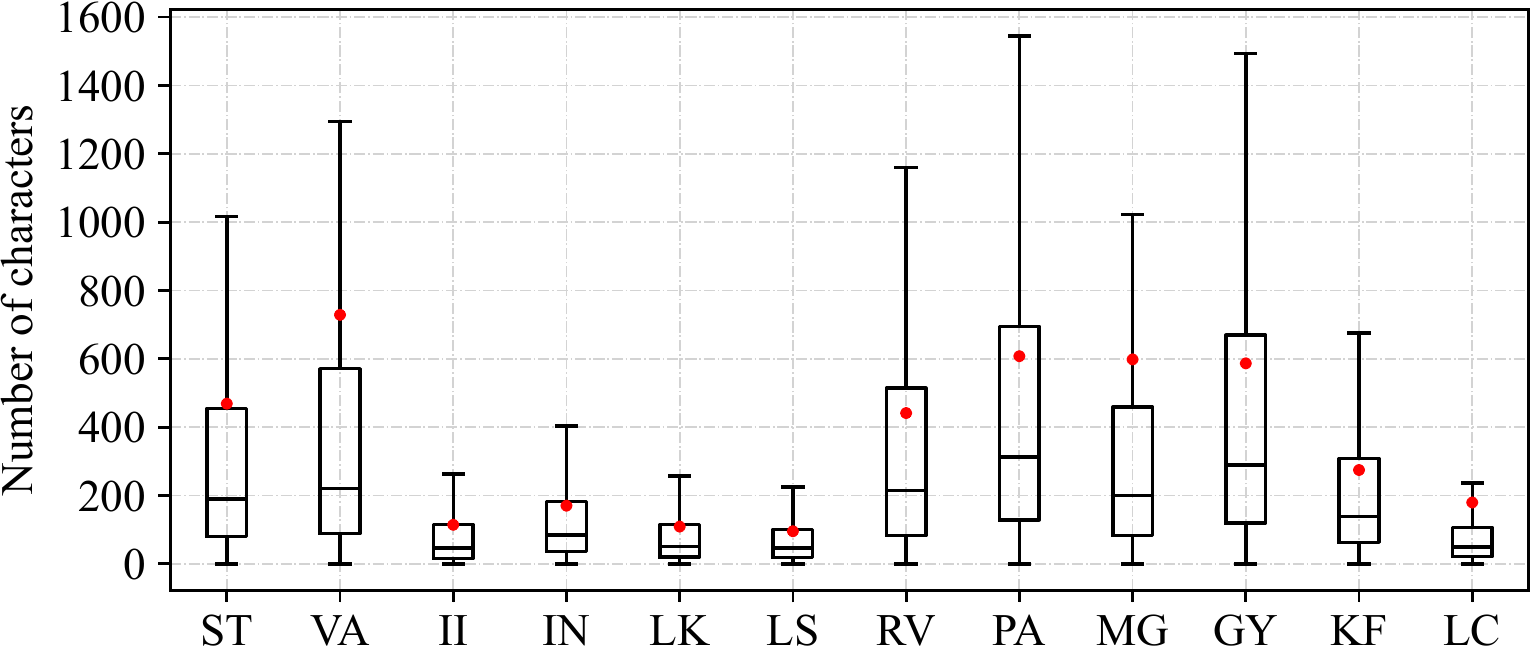}
    \caption{Post length distribution of each forum. Red dots are means. Abbreviations are shown in Table~\ref{tab:database-taxonomy}.} 
    \label{fig:post-length}
    \vspace{3.5mm}
    \centering
    \includegraphics[width=.47\textwidth]{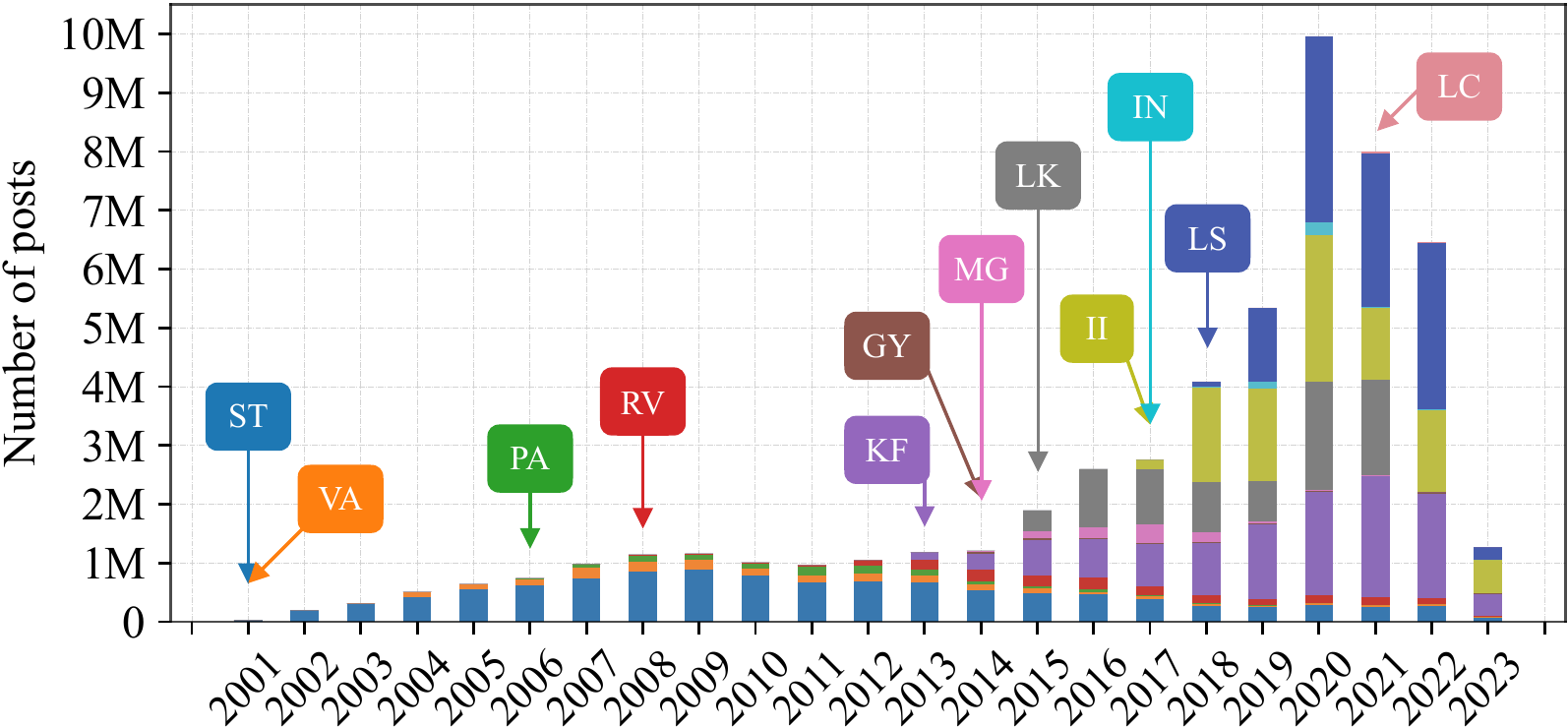}
    \caption{Number of forum posts; coloured labels mark forum inceptions. Abbreviations are shown in Table~\ref{tab:database-taxonomy}.} 
    \label{fig:timeline-overview}\vspace{-2mm}
\end{figure}

\para{\pacat} This includes two popular men's online forums offering instructions, advice and training to pick up, date, and have sex with women. The so-called `coaches' on the forums share their experiences and discuss techniques for bullying and tricking women into sex and relationships; many of these techniques are seen by many women as sexual harassment~\cite{bates2021men}. Forum users exhibit toxic, sexist ideas and violence against women~\cite{banet2019pick}.

\para{\mmcat} This includes two influential forums of movements whose goal is to completely separate men from women, protecting and preserving male sovereignty while exhibiting antifeminist, misogynistic and toxic-masculine ideals.

\para{\tdcat} This includes the two largest online harassment forums associated with real-world attacks on transgender, feminist, autistic and disabled individuals, with members engaging in organised trolling, cyberstalking, doxxing and swatting of their targets, some of whom have committed suicide after being harassed~\cite{ambreen2019fyi}.

\begin{table*}[t]
\centering
\caption{User overlaps across forums. $n(i,j)$ depicts the number of users on forum $i$ who also register on forum $j$.}
\small
\label{tab:user-overlap}
\setlength{\tabcolsep}{1.1em}
\begin{tabular}{l|rrrrrrrrrrrr}
\midrule
$n$ & \stabbr & \vaabbr & \iiabbr & \inabbr & \lkabbr & \lsabbr & \rvabbr & \paabbr & \mgabbr & \gyabbr & \kfabbr & \lcabbr\\
\midrule
\stabbr & -- & 945 & 403 & 154 & 673 & 412 & 906 & 2\,346 & 273 & 149 & 1\,624 & 14  \\
\vaabbr & 945 & -- & 56 & 27 & 86 & 48 & 109 & 203 & 35 & 20 & 166 & 0  \\
\iiabbr & 403 & 56 & -- & 205 & 549 & 769 & 144 & 297 & 44 & 31 & 365 & 4  \\
\inabbr & 154 & 27 & 205 & -- & 127 & 116 & 45 & 119 & 27 & 18 & 139 & 2  \\
\lkabbr & 673 & 86 & 549 & 127 & -- & 993 & 268 & 596 & 79 & 52 & 546 & 8  \\
\lsabbr & 412 & 48 & 769 & 116 & 993 & -- & 155 & 360 & 62 & 23 & 299 & 2  \\
\rvabbr & 906 & 109 & 144 & 45 & 268 & 155 & -- & 815 & 117 & 79 & 477 & 5  \\
\paabbr & 2\,346 & 203 & 297 & 119 & 596 & 360 & 815 & -- & 212 & 114 & 1\,113 & 16  \\
\mgabbr & 273 & 35 & 44 & 27 & 79 & 62 & 117 & 212 & -- & 120 & 158 & 1  \\
\gyabbr & 149 & 20 & 31 & 18 & 52 & 23 & 79 & 114 & 120 & -- & 84 & 0  \\
\kfabbr & 1\,624 & 166 & 365 & 139 & 546 & 299 & 477 & 1\,113 & 158 & 84 & -- & 97  \\
\lcabbr & 14 & 0 & 4 & 2 & 8 & 2 & 5 & 16 & 1 & 0 & 97 & --  \\
\midrule
\end{tabular}
\end{table*}

\begin{figure*}[t]
    \centering
    \includegraphics[width=\textwidth]{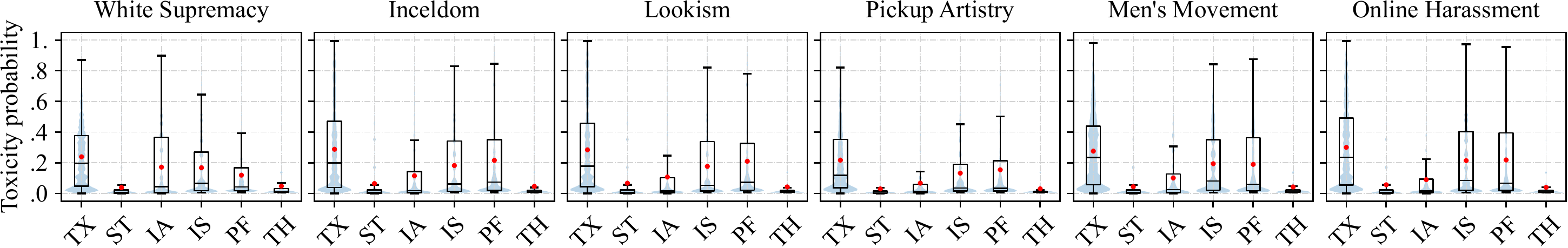}
    \caption{The toxicity level of posts. The purple areas: probability density; TX: toxicity, ST: severe toxicity, IA: identity attack, IS: insult, PF: profanity, TH: threat. Red dots are means. A large proportion of posts are not toxic.}
    \label{fig:toxicity-distribution}
    \vspace{3.5mm}
    \includegraphics[width=0.16\textwidth]{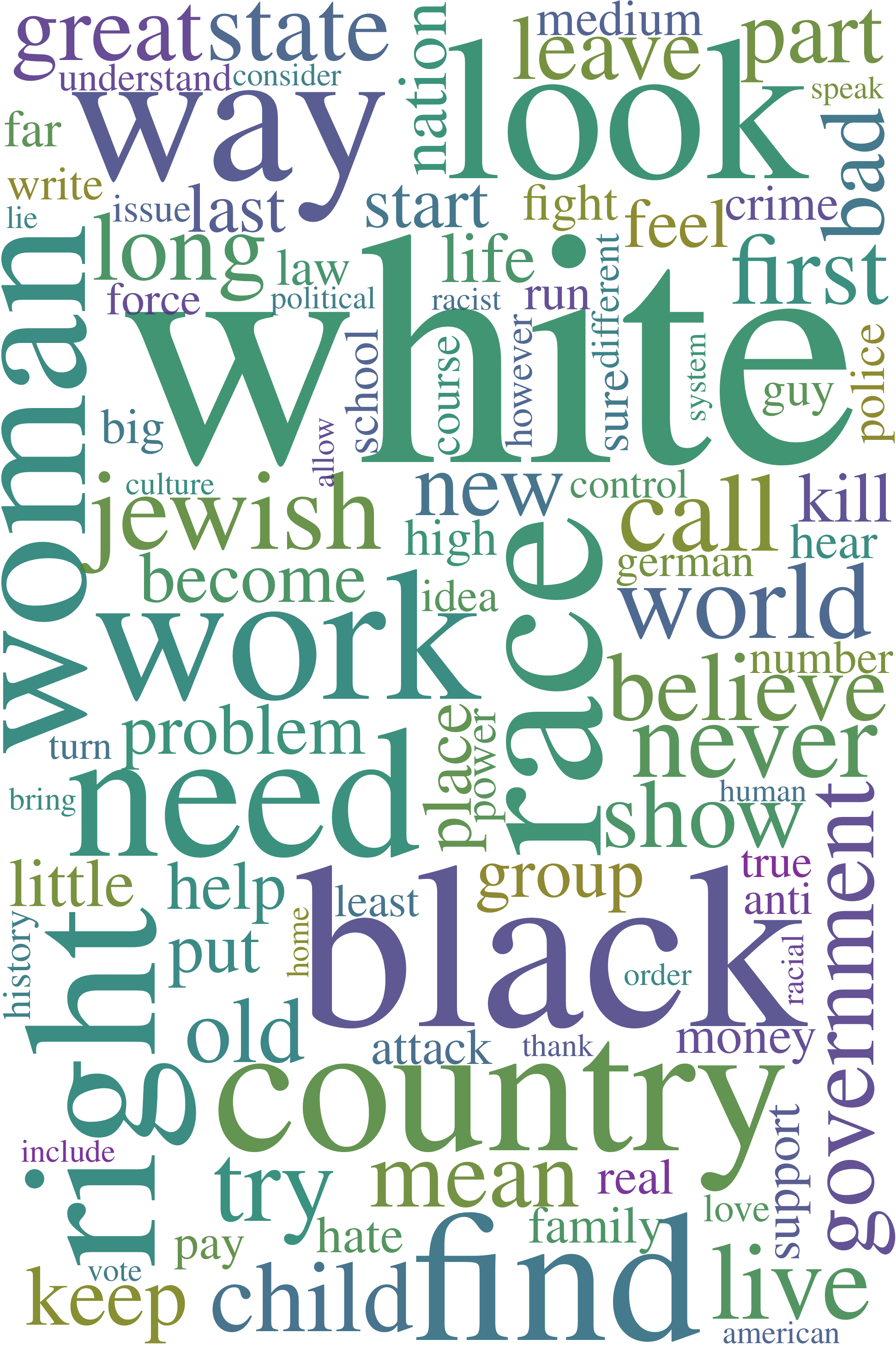}\hspace{0.00225\textwidth}
    \includegraphics[width=0.16\textwidth]{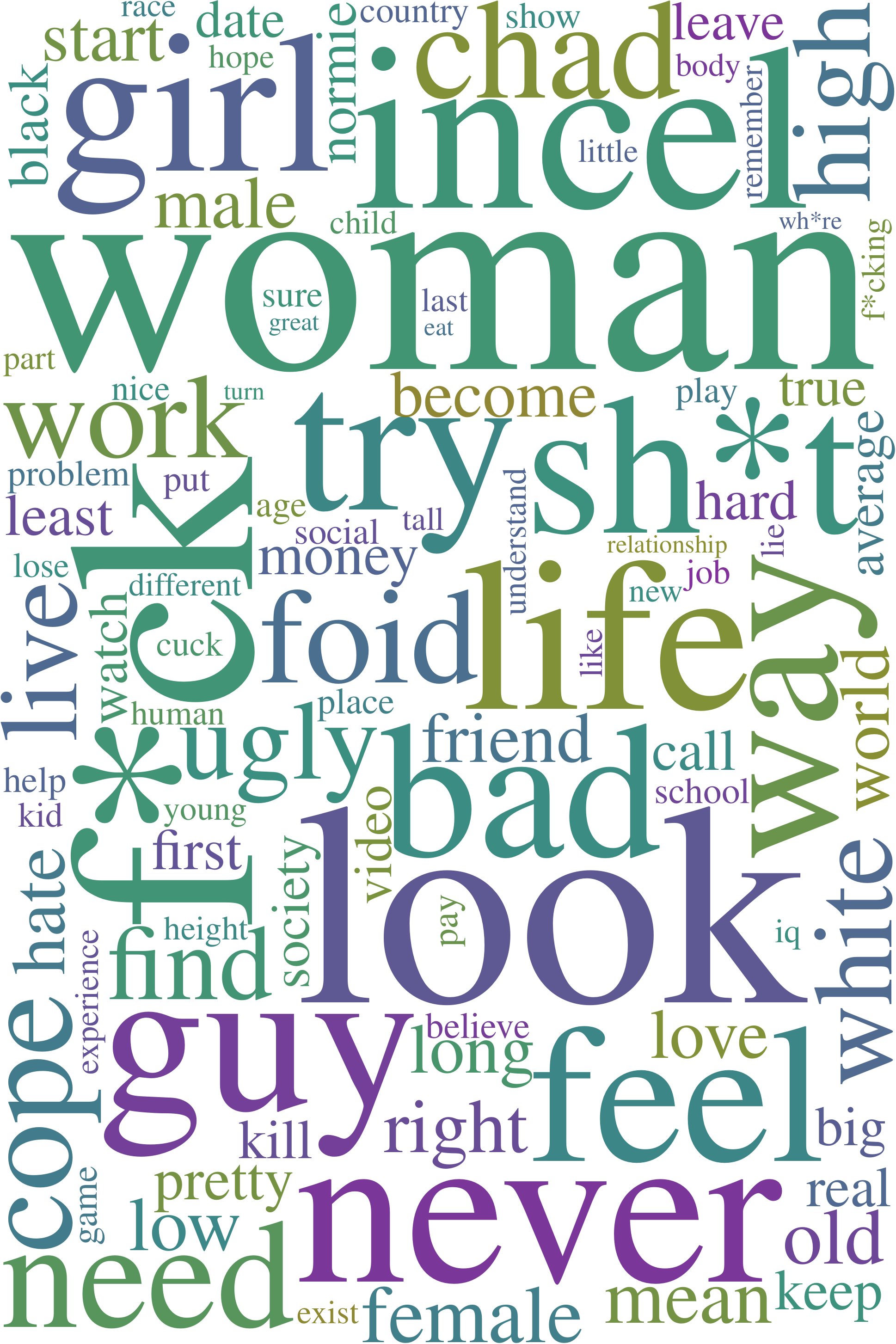}\hspace{0.00225\textwidth}
    \includegraphics[width=0.16\textwidth]{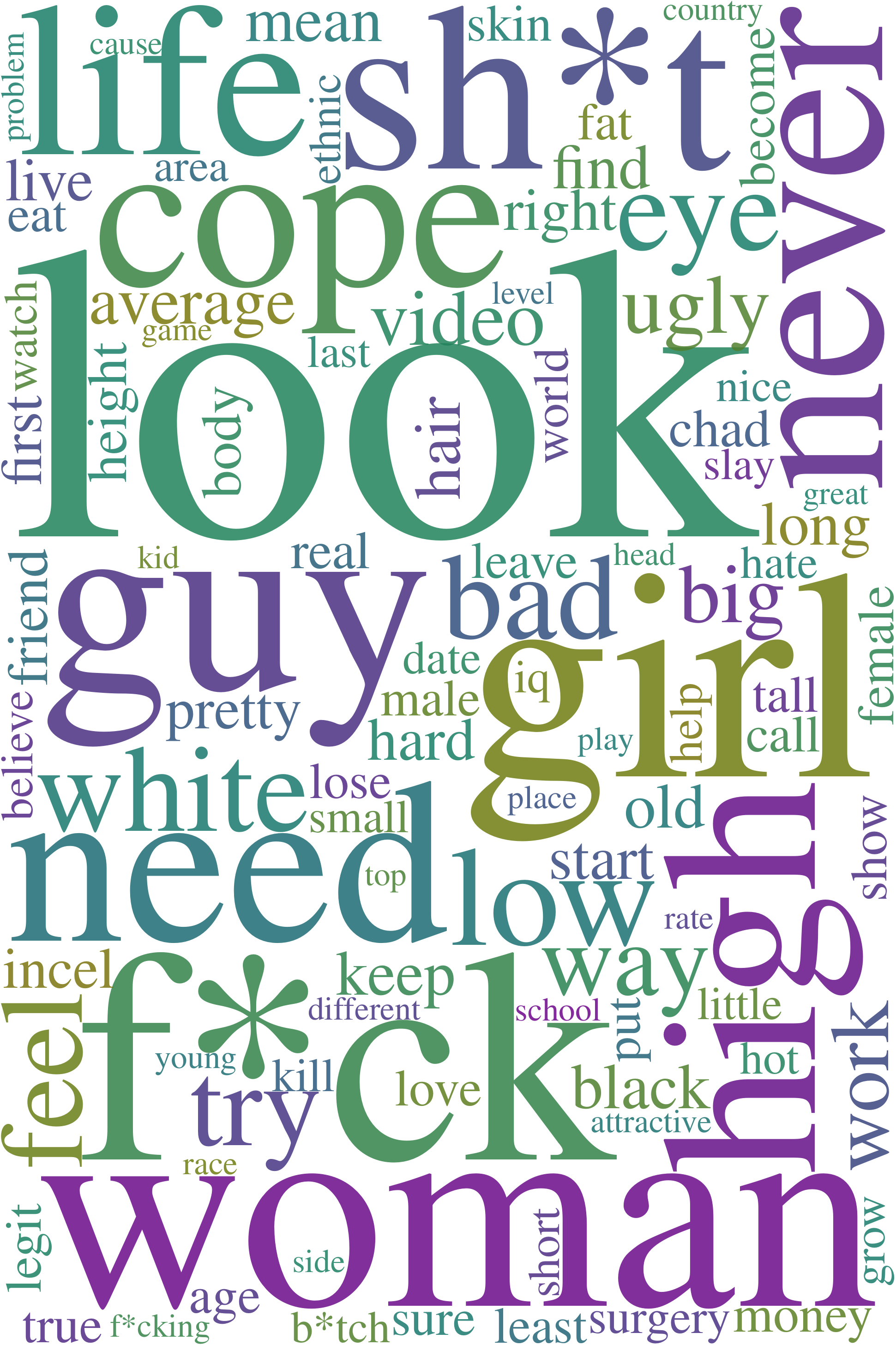}\hspace{0.00225\textwidth}
    \includegraphics[width=0.16\textwidth]{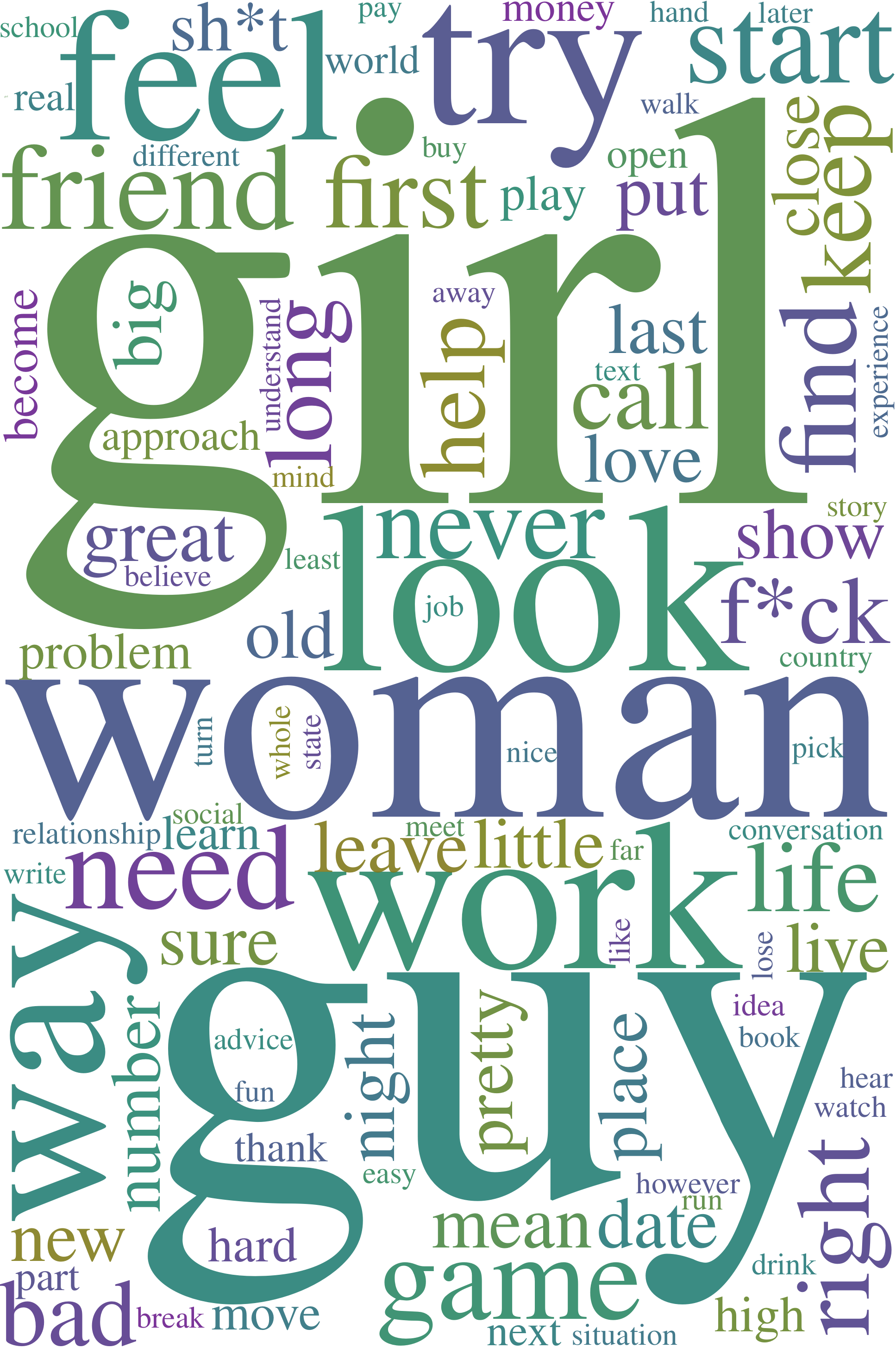}\hspace{0.00225\textwidth}
    \includegraphics[width=0.16\textwidth]{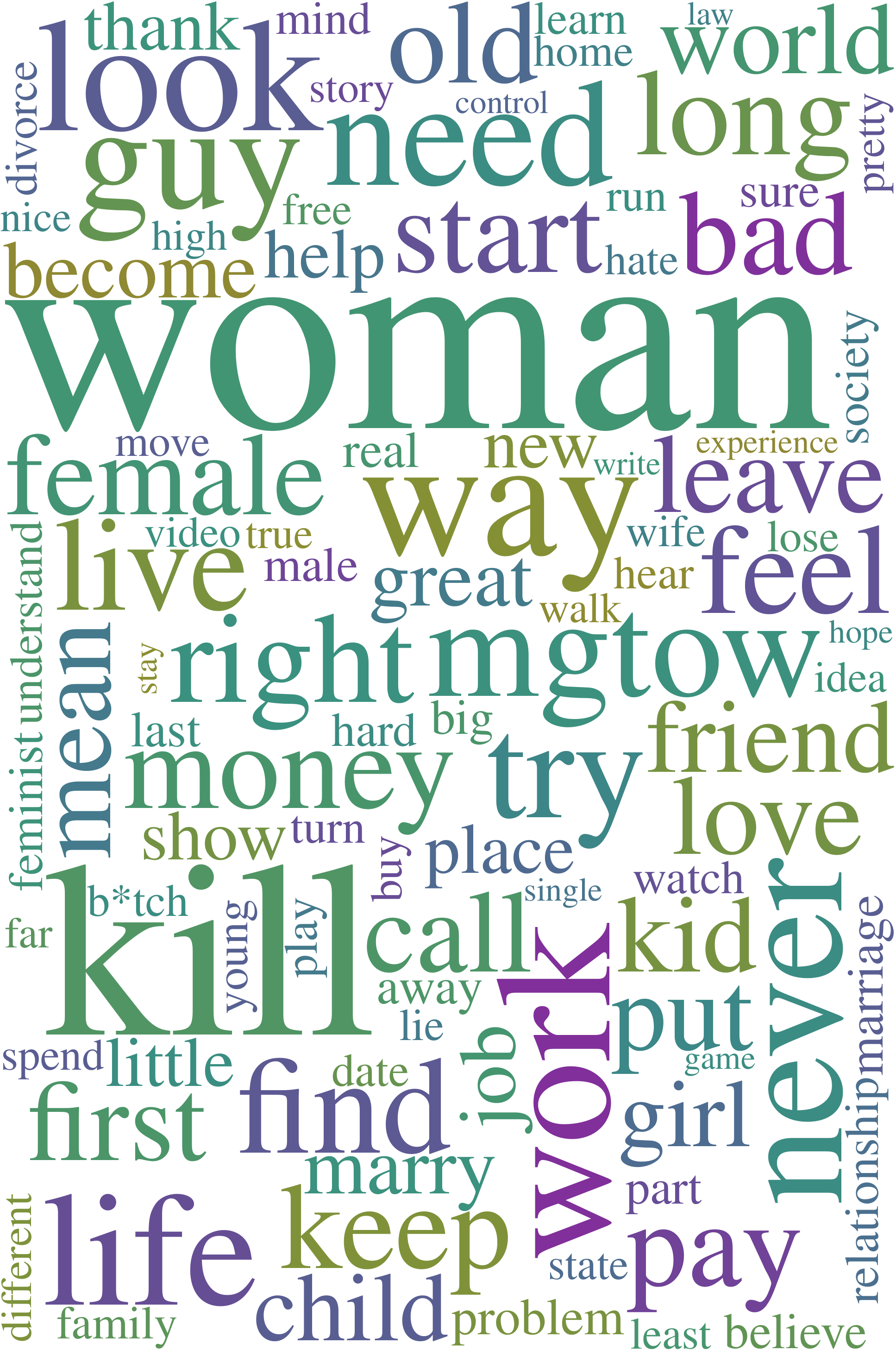}\hspace{0.00225\textwidth}
    \includegraphics[width=0.16\textwidth]{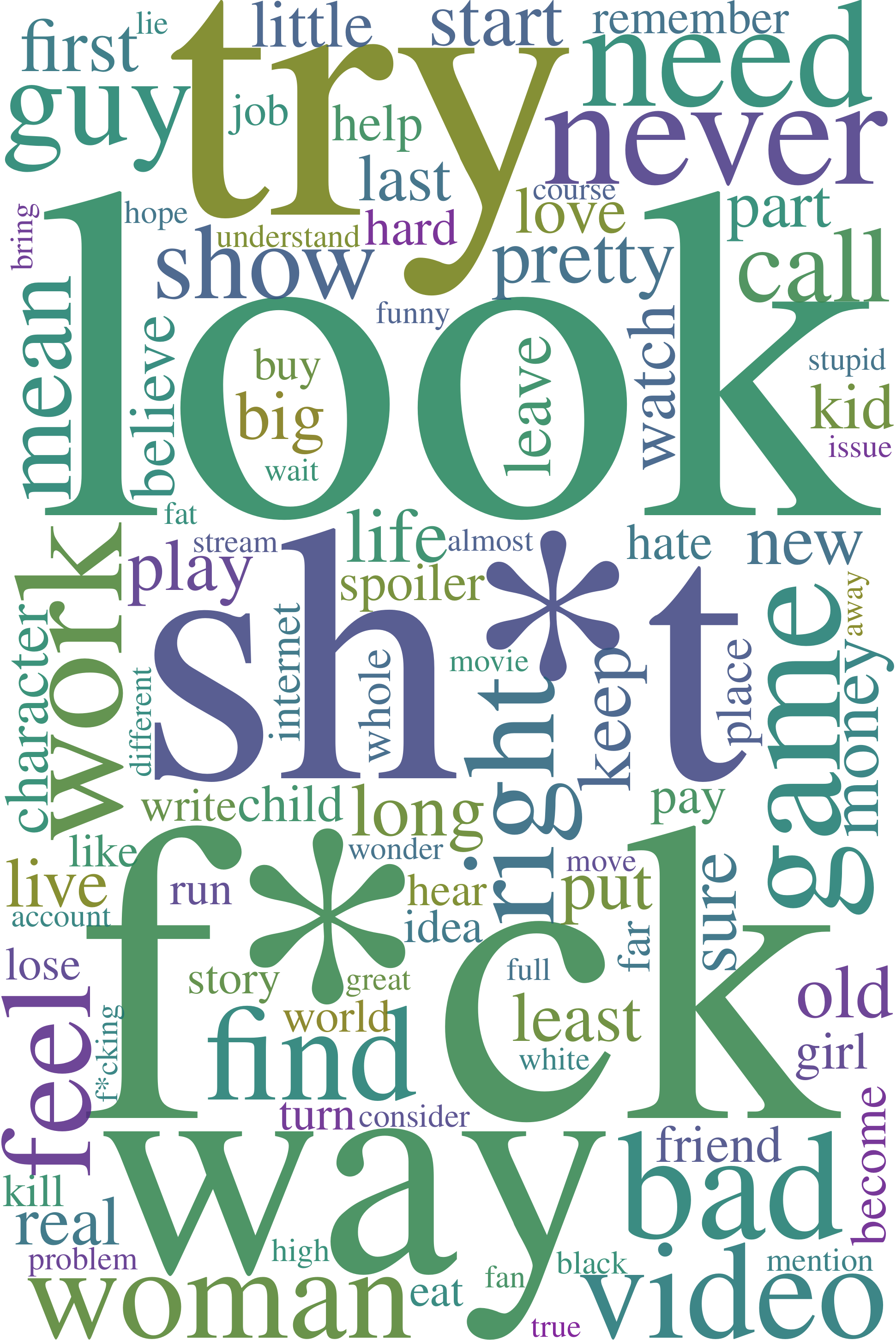}
    \caption{Word clouds of the top 100 keywords on \wscat, \iccat, \lkcat, \pacat, \mmcat, and \tdcat~forums, respectively. The word sizes correspond to their popularity.}
    \label{fig:word-cloud}\vspace{-2mm}
\end{figure*}

\section{The \extremebb Database} \label{sec:dataset}
\extremebb has over \nPostsTotalShorten~posts (Table~\ref{tab:database-taxonomy}); the post length distribution is shown in Figure~\ref{fig:post-length}. Just as with cybercrime forums~\cite{vu2020turning}, increasing activity was seen on extremist forums during the \covid pandemic, with posting in 2020 being double that of 2019 (but decreasing 20\% in 2021), presumably as users had more time online during lockdowns (Figure~\ref{fig:timeline-overview}). Thousands of users join multiple forums,\footnote{~A lower-bound overlap is estimated based on identical rare usernames and correlated posting patterns. Cross-forum analysis may be limited due to the lack of ground-truth insight.} suggesting considerable overlaps between different types of online extremism (see Table~\ref{tab:user-overlap}). These forums are considered extreme, but Google's Perspective API rates a large proportion of individual posts as non-toxic, when measured in January 2023 (Figure~\ref{fig:toxicity-distribution}).\footnote{~This is the state-of-the-art tool for toxicity analysis~\cite{zannettou2020measuring}, but determining posts' toxicity may heavily depend on context. The tool only looks at the first 501 characters of each post~\cite{anna2023hacker}, and is limited in dealing with both adversarial samples~\cite{hosseini2017deceiving} and racial bias~\cite{sap2019risk}.} The main themes of each category are shown in Figure~\ref{fig:word-cloud}, with stop words excluded and terms lemmatised then obfuscated~\cite{nozza-hovy-2023-prof}.

\para{Data Collection}\label{subsec:method-data-collection} 
\extremebb is maintained and expanded continuously, and in real time when necessary. Our web scraper, based on CrimeBot~\cite{sergiocrimebb}, can bypass various types of \captcha and other bot prevention techniques. It is designed to ensure data completeness by visiting each board regularly while persistently storing its progress for incremental crawling when disruptions occur. A new crawl starts when one finishes: some sites complete within hours, while some take days or weeks. Vanished valuable forums such as \lookism and \lolcow have been fully captured.

\para{Data Licensing} Our regime was carefully crafted in consultation with legal academics, university lawyers and specialist external counsel, allowing the data to be shared in multiple jurisdictions. \extremebb aims to allow both computer and social scientists -- especially those who may lack time or technical skills for data collection -- to focus on their research instead of spending months or even years gathering the data. Although the forums are public, some content may be personally identifiable, or even illegal. Licensees are therefore required to sign a formal agreement with the \href{https://www.cambridgecybercrime.uk/}{\ccc} to prevent misuse. \extremebb is provided as PostgreSQL snapshots; once registered, licensees will automatically get access to the most recent data as they are published. Shared data can only be used for research. Data leakage is a contract breach regardless of whether directly exposing the data or carelessly publishing a model from which the data could be extracted.

\para{Limitations} Our database is unlikely to capture all available extremist forums, but we believe it includes the most prominent ones. Our ethical safeguards do not allow automatic downloads of images because of the risk of potentially illegal material, with the result that \nEmptyPostsPropsTotal\% of posts are empty. The database thus includes text only, but media URLs are kept for further analyses if needed. Some forums have fewer posts than the statistics shown on their homepage as (1) our collection started in 2020 thus older content might have been removed or hidden, (2) content posted then removed quickly between crawls may be missed, and (3) access to some sub-forums may be restricted. \extremebb only consists of English forums, but we are also starting to expand it to other languages.

\para{Our Vision} We aim to offer a high-quality, fine-grained dataset to enable researchers to better understand the behaviours and interactions of extremists within and across different subcultures, and to monitor incidents affecting them such as law-enforcement interventions, the pandemic~\cite{vu2020pandemic} and platform disruption~\cite{vu2023no}. Our data are mostly unlabelled and can best be used for statistical or unsupervised tasks, but we have started labelling hate speech for language-model training and evaluation. We have built an interactive web-based tool to assist less technical scientists to explore our cybercrime data~\cite{pete2022postcog}; this will shortly be extended to \extremebb.

\section{Ethics Statement} Our team has robust ethical procedures to deal with scraped data that may contain sensitive information. The collection of \extremebb and the license agreement were approved by our institutional ethics committee. We only collect publicly accessible posts, which solves many but not all problems of privacy~\cite{webscrapinglegal}. Our scraper does not flood targeted servers with unnecessary traffic; we do our best to collect only what is needed. The scraper operates slowly and stores raw HTML files to avoid repetitive crawls in case offline parsing is required. The parsed data are kept encrypted to avoid data leakage. Carrying out research on sensitive and abusive data may introduce a wide range of risks for researchers~\cite{doerfler2021m}, so we follow guidance to protect researchers from online harassment~\cite{marwick2016best}. One licensing condition is that every project must be approved by the licensee's institutional ethics committee or review board; another is that research results must be placed in the public domain, ensuring transparency. We only analyse aggregated behaviours of groups instead of studying individuals, which follows the British Society of Criminology Statement on Ethics~\cite{britishethics}. 

\section*{Acknowledgments}
We are grateful to Alice Hutchings, Ben Collier, and Sergio Pastrana for their insightful feedback, as well as Richard Clayton and Tina Marjanov for maintaining the data scrapers' infrastructure. This work was supported by the Engineering and Physical Sciences Research Council (EPSRC) [grant number EP/V026178/1] and the European Research Council (ERC) under the European Union’s Horizon 2020 research and innovation programme [grant agreement No 949127].

\bibliography{main}
\bibliographystyle{acl2023}
\end{document}

%% file: configs.tex
\newcommand{\covid}{{\small \scshape Covid-19}\xspace}
\newcommand{\extremebb}{{ExtremeBB}\xspace}
\newcommand{\extremebbnosize}{{\scshape ExtremeBB}\xspace}
\newcommand{\captcha}{{\scshape Captchas}\xspace}
\newcommand{\anh}[1]{\{{\textcolor{blue}{anh: #1}}\}}

\newcommand{\para}[1]{\noindent\textbf{#1.}}
\newcommand{\wscat}{{\scshape \small White Supremacy}\xspace}
\newcommand{\lkcat}{{\scshape \small Lookism}\xspace}
\newcommand{\iccat}{{\scshape \small Inceldom}\xspace}
\newcommand{\pacat}{{\scshape \small Pickup Artistry}\xspace}
\newcommand{\mmcat}{{\scshape \small Men's Movement}\xspace}
\newcommand{\tdcat}{{\scshape \small Online Harassment}\xspace}

\newcommand{\wscatnosize}{{\scshape White Supremacy}\xspace}
\newcommand{\lkcatnosize}{{\scshape Lookism}\xspace}
\newcommand{\iccatnosize}{{\scshape Inceldom}\xspace}
\newcommand{\pacatnosize}{{\scshape Pickup Artistry}\xspace}
\newcommand{\mmcatnosize}{{\scshape Men's Movement}\xspace}
\newcommand{\tdcatnosize}{{\scshape Trolling} \& {\scshape Doxxing}\xspace}

\newcommand{\stormfront}{{\scshape \small  Stormfront}\xspace}
\newcommand{\vnn}{{\scshape \small Vanguard News Network}\xspace}
\newcommand{\incelsis}{{\scshape \small Incelsis}\xspace}
\newcommand{\incelsnet}{{\scshape \small Incelsnet}\xspace}
\newcommand{\looksmax}{{\scshape \small Looksmax}\xspace}
\newcommand{\lookism}{{\scshape \small Lookism}\xspace}
\newcommand{\pa}{{\scshape \small Pickup Artist}\xspace}
\newcommand{\rooshv}{{\scshape \small Roosh V}\xspace}
\newcommand{\kiwifarms}{{\scshape \small Kiwi Farms}\xspace}
\newcommand{\mgtow}{{\scshape \small Men Going Their Own Way}\xspace}
\newcommand{\gyow}{{\scshape \small Going Your Own Way}\xspace}
\newcommand{\lolcow}{{\scshape \small Lolcow}\xspace}
\newcommand{\lookstheory}{{\scshape \small Looks Theory}\xspace}
\newcommand{\chimpout}{{\scshape \small Chimpout}\xspace}

\newcommand{\stormfrontnosize}{{\scshape Stormfront}\xspace}
\newcommand{\vnnnosize}{{\scshape Vanguard News Network}\xspace}
\newcommand{\incelsisnosize}{{\scshape Incelsis}\xspace}
\newcommand{\incelsnetnosize}{{\scshape Incelsnet}\xspace}
\newcommand{\looksmaxnosize}{{\scshape Looksmax}\xspace}
\newcommand{\lookismnosize}{{\scshape Lookism}\xspace}
\newcommand{\panosize}{{\scshape Pickup Artist}\xspace}
\newcommand{\rooshvnosize}{{\scshape Roosh V}\xspace}
\newcommand{\kiwifarmsnosize}{{\scshape Kiwi Farms}\xspace}
\newcommand{\mgtownosize}{{\scshape Men Going Their Own Way}\xspace}
\newcommand{\gyownosize}{{\scshape Going Your Own Way}\xspace}
\newcommand{\lookstheorynosize}{{\scshape Looks Theory}\xspace}
\newcommand{\chimpoutnosize}{{\scshape Chimpout}\xspace}
\newcommand{\lolcownosize}{{\scshape Lolcow}\xspace}

\newcommand{\stormfronthomepage}{\href{http://stormfront.org}{https://stormfront.org}}
\newcommand{\vnnhomepage}{\href{http://vnnforum.com}{https://vnnforum.com}}
\newcommand{\incelsishomepage}{\href{http://incels.is}{https://incels.is}}
\newcommand{\incelsnethomepage}{\href{http://incels.net}{https://incels.net}}
\newcommand{\looksmaxhomepage}{\href{http://looksmax.org}{https://looksmax.org}}
\newcommand{\lookismhomepage}{\href{http://lookism.net}{https://lookism.net}}
\newcommand{\pahomepage}{\href{http://pick-up-artist-forum.com}{https://pick-up-artist-forum.com}}
\newcommand{\rooshvhomepage}{\href{http://rooshvforum.com}{https://rooshvforum.com}}
\newcommand{\kiwifarmshomepage}{\href{http://kiwifarms.net}{https://kiwifarms.net}}
\newcommand{\mgtowhomepage}{\href{http://mgtow.com/forums}{https://mgtow.com/forums}}
\newcommand{\gyowhomepage}{\href{http://goingyourownway.com}{https://goingyourownway.com}}
\newcommand{\lolcowhomepage}{\href{http://lolcow.org}{https://lolcow.org}}

\newcommand{\stabbr}{{\scshape \small ST}\xspace}
\newcommand{\vaabbr}{{\scshape \small VA}\xspace}
\newcommand{\iiabbr}{{\scshape \small II}\xspace}
\newcommand{\inabbr}{{\scshape \small IN}\xspace}
\newcommand{\lsabbr}{{\scshape \small LS}\xspace}
\newcommand{\lkabbr}{{\scshape \small LK}\xspace}
\newcommand{\paabbr}{{\scshape \small PA}\xspace}
\newcommand{\rvabbr}{{\scshape \small RV}\xspace}
\newcommand{\kfabbr}{{\scshape \small KF}\xspace}
\newcommand{\lcabbr}{{\scshape \small LC}\xspace}
\newcommand{\mgabbr}{{\scshape \small MG}\xspace}
\newcommand{\gyabbr}{{\scshape \small GY}\xspace}

\newcommand{\whitenations}{{\scshape \small White Nations}\xspace}
\newcommand{\ironvolk}{{\scshape \small Iron Volk}\xspace}
\newcommand{\christogenea}{{\scshape \small Christogenea}\xspace}
\newcommand{\creativityalliance}{{\scshape \small Creativity Alliance}\xspace}

\newcommand{\whitenationsnosize}{{\scshape White Nations}\xspace}
\newcommand{\ironvolknosize}{{\scshape Iron Volk}\xspace}
\newcommand{\christogeneanosize}{{\scshape Christogenea}\xspace}
\newcommand{\creativityalliancenosize}{{\scshape Creativity Alliance}\xspace}

\newcommand{\ccc}{Cambridge Cybercrime Centre}

%% file: stats.tex
\newcommand{\nUsersStormfront}{{137\,292}}
\newcommand{\nBoardsStormfront}{{152}}
\newcommand{\nThreadsStormfront}{{782\,347}}
\newcommand{\nPostsStormfront}{{10\,703\,957}}
\newcommand{\nEmptyPostsPropsStormfront}{{2.65}}
\newcommand{\nDataPeriodStormfront}{{2001/08-2023/04}}
\newcommand{\nUsersVanguardNewsNetwork}{{8\,504}}
\newcommand{\nBoardsVanguardNewsNetwork}{{108}}
\newcommand{\nThreadsVanguardNewsNetwork}{{206\,300}}
\newcommand{\nPostsVanguardNewsNetwork}{{1\,712\,924}}
\newcommand{\nEmptyPostsPropsVanguardNewsNetwork}{{1.65}}
\newcommand{\nDataPeriodVanguardNewsNetwork}{{2001/10-2023/04}}
\newcommand{\nUsersIncelsis}{{13\,825}}
\newcommand{\nBoardsIncelsis}{{7}}
\newcommand{\nThreadsIncelsis}{{439\,675}}
\newcommand{\nPostsIncelsis}{{9\,030\,678}}
\newcommand{\nEmptyPostsPropsIncelsis}{{11.34}}
\newcommand{\nDataPeriodIncelsis}{{2017/11-2023/04}}
\newcommand{\nUsersIncelsnet}{{3\,614}}
\newcommand{\nBoardsIncelsnet}{{6}}
\newcommand{\nThreadsIncelsnet}{{23\,110}}
\newcommand{\nPostsIncelsnet}{{364\,183}}
\newcommand{\nEmptyPostsPropsIncelsnet}{{3.87}}
\newcommand{\nDataPeriodIncelsnet}{{2017/11-2022/11}}
\newcommand{\nUsersLookism}{{16\,362}}
\newcommand{\nBoardsLookism}{{10}}
\newcommand{\nThreadsLookism}{{770\,429}}
\newcommand{\nPostsLookism}{{7\,294\,617}}
\newcommand{\nEmptyPostsPropsLookism}{{8.65}}
\newcommand{\nDataPeriodLookism}{{2015/06-2021/12}}
\newcommand{\nUsersLooksmax}{{12\,570}}
\newcommand{\nBoardsLooksmax}{{17}}
\newcommand{\nThreadsLooksmax}{{650\,126}}
\newcommand{\nPostsLooksmax}{{10\,177\,131}}
\newcommand{\nEmptyPostsPropsLooksmax}{{7.21}}
\newcommand{\nDataPeriodLooksmax}{{2018/08-2023/04}}
\newcommand{\nUsersRooshV}{{13\,389}}
\newcommand{\nBoardsRooshV}{{33}}
\newcommand{\nThreadsRooshV}{{36\,849}}
\newcommand{\nPostsRooshV}{{1\,656\,740}}
\newcommand{\nEmptyPostsPropsRooshV}{{4.81}}
\newcommand{\nDataPeriodRooshV}{{2008/08-2023/04}}
\newcommand{\nUsersPickupArtist}{{65\,115}}
\newcommand{\nBoardsPickupArtist}{{43}}
\newcommand{\nThreadsPickupArtist}{{177\,438}}
\newcommand{\nPostsPickupArtist}{{943\,853}}
\newcommand{\nEmptyPostsPropsPickupArtist}{{0.22}}
\newcommand{\nDataPeriodPickupArtist}{{2006/03-2023/04}}
\newcommand{\nUsersMenGoingTheirOwnWay}{{3\,870}}
\newcommand{\nBoardsMenGoingTheirOwnWay}{{18}}
\newcommand{\nThreadsMenGoingTheirOwnWay}{{55\,093}}
\newcommand{\nPostsMenGoingTheirOwnWay}{{853\,286}}
\newcommand{\nEmptyPostsPropsMenGoingTheirOwnWay}{{2.02}}
\newcommand{\nDataPeriodMenGoingTheirOwnWay}{{2014/07-2021/05}}
\newcommand{\nUsersGoingYourOwnWay}{{1\,974}}
\newcommand{\nBoardsGoingYourOwnWay}{{21}}
\newcommand{\nThreadsGoingYourOwnWay}{{15\,578}}
\newcommand{\nPostsGoingYourOwnWay}{{189\,862}}
\newcommand{\nEmptyPostsPropsGoingYourOwnWay}{{1.00}}
\newcommand{\nDataPeriodGoingYourOwnWay}{{2014/02-2023/04}}
\newcommand{\nUsersKiwiFarms}{{61\,432}}
\newcommand{\nBoardsKiwiFarms}{{51}}
\newcommand{\nThreadsKiwiFarms}{{51\,337}}
\newcommand{\nPostsKiwiFarms}{{10\,528\,403}}
\newcommand{\nEmptyPostsPropsKiwiFarms}{{3.51}}
\newcommand{\nDataPeriodKiwiFarms}{{2013/01-2023/04}}
\newcommand{\nUsersLolcow}{{543}}
\newcommand{\nBoardsLolcow}{{19}}
\newcommand{\nThreadsLolcow}{{2\,042}}
\newcommand{\nPostsLolcow}{{56\,990}}
\newcommand{\nEmptyPostsPropsLolcow}{{6.96}}
\newcommand{\nDataPeriodLolcow}{{2021/04-2022/06}}
\newcommand{\nUsersTotal}{{338\,490}}
\newcommand{\nBoardsTotal}{{485}}
\newcommand{\nThreadsTotal}{{3\,210\,324}}
\newcommand{\nPostsTotal}{{53\,512\,624}}
\newcommand{\nEmptyPostsPropsTotal}{{5.96}}
\newcommand{\nDataPeriodTotal}{{2001/08-2023/04}}
\newcommand{\nUsersTotalShorten}{{338.5k}}
\newcommand{\nThreadsTotalShorten}{{3.2M}}
\newcommand{\nPostsTotalShorten}{{53.5M}}